\documentclass[a4paper,11pt]{article}
\usepackage{pos}
\usepackage{wrapfig}
\usepackage{graphicx}

\newlength{\bibitemsep}
\newlength{\bibparskip}
\let\oldthebibliography\thebibliography
\renewcommand\thebibliography[1]{%
  \oldthebibliography{#1}%
  \setlength{\parskip}{\bibitemsep}%
  \setlength{\itemsep}{\bibparskip}%
}

\newcommand{\lsim}{\mathrel{\hbox{\rlap{\lower.75ex \hbox{$\sim$}} \kern-.3em \raise.4ex \hbox{$<$}}}}
\newcommand{\gsim}{\mathrel{\hbox{\rlap{\lower.75ex \hbox{$\sim$}} \kern-.3em \raise.4ex \hbox{$>$}}}}

\title{The Silicon Strip Detector Subsystem for the Trans-Iron Galactic Element Recorder for the International Space Station (TIGERISS)}
\ShortTitle{TIGERISS SSD System}

\author[a]{John F. Krizmanic}
\author*[b]{Scott Nutter}

\affiliation[a]{Astroparticle Physics Laboratory, NASA Goddard Space Flight Center\\
Greenbelt, Maryland 20771 USA}

\affiliation[b]{Department of Physics and Astronomy, Northern Kentucky University\\
 Highland Heights, KY  USA}

\onbehalf{for the TIGERISS Collaboration} 

\emailAdd{john.f.krizmanic@nasa.gov}
\emailAdd{nutters@nku.edu}

\abstract{The Trans-Iron Galactic Element Recorder for the International Space Station (TIGERISS) is under construction and is planned for launch in 2027 and will be  attached at the SOX location on the Columbus module on the ISS. TIGERISS will make the first definitive measurements of Ultra-Heavy Galactic Cosmic Rays (UHGCRs; Z >29) on an individual element basis past barium ($^{56}$Ba), through the lanthinides, and to lead ($^{82}$Pb). TIGERISS has a geometry factor of 1.06 m$^2$ sr and is comprised of four planes of single-sided silicon strip detectors (SSDs) arranged in orthogonal X-Y layers with an X-Y pair above and an X-Y pair below two large-area Cherenkov detectors. The top Cherenkov detector is comprised of a mosaic of aerogel radiators (n =1.05) while the bottom Cherenkov detector has an acrylic radiator (n = 1.49). The combination of the Cherenkov velocity measurements with the precise measurements of the ionization and trajectory of the traversing cosmic rays leads to highly accurate charge measurements of $<$ 0.25 c.u. over the entire elemental range of $^5$B through $^{82}$Pb. These TIGERISS measurements are highly sensitive in determining the strength of s-process, r-process, and rp-processes of Galactic nucleosynthesis while providing critical data needed for multi-messenger studies to determine the contributions of extreme phenomena, including supernovae (SN) and Neutron Star Mergers (NSMs), in the production of galactic matter. The science goals of TIGERISS, mission status, instrument design and performance of the TIGERISS SSD subsystem in relation to the measurements and science goals of TIGERISS are discussed in this paper.
}

\ConferenceLogo{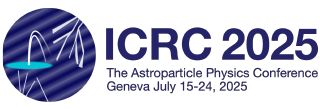}

\FullConference{39th International Cosmic Ray Conference (ICRC2025)\\
 15–24 July 2025\\
Geneva, Switzerland\\}

\begin{document}
\maketitle

\begin{figure}[!t]
    \centering
    \includegraphics[width=\textwidth]{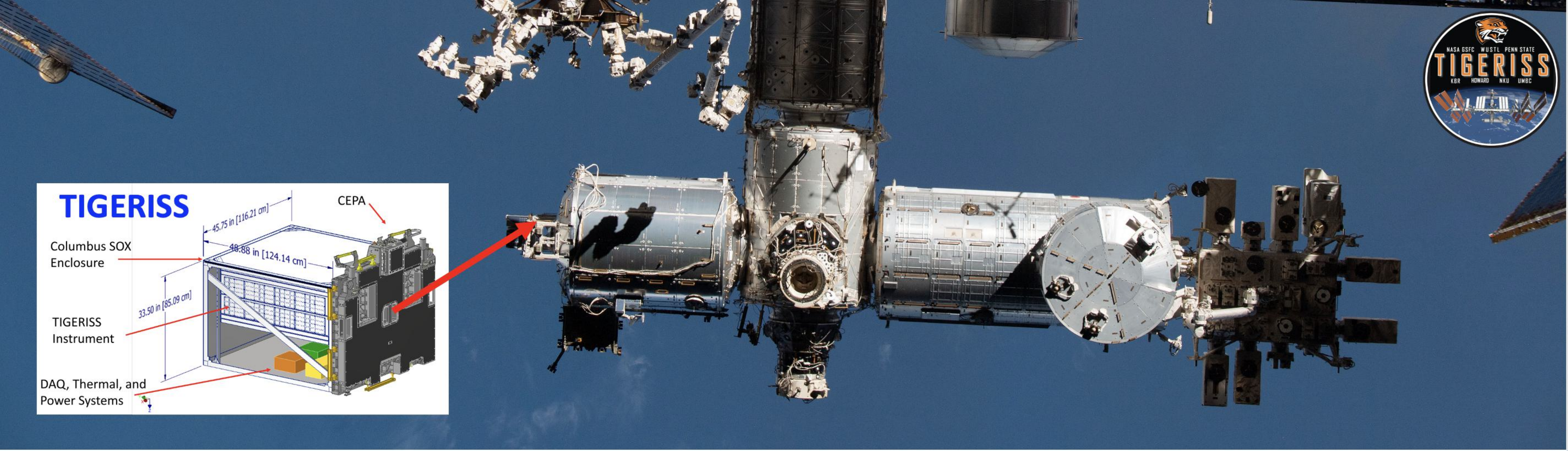}
    \caption{The Columbus module on the International Space Station. TIGERISS is manifested to be attached to the SOX external payload attachment point in 2027.}
    \label{fig:TIGERISS}
\end{figure}

\section{Introduction}

The Trans-Iron Galactic Element Recorder for the International Space Station (TIGERISS) experiment is a NASA-funded Pioneers-class mission designed to make accurate measurements of the individual element abundances from boron ($^5$B) thru lead ($^{82}$Pb) in the cosmic radiation.  Thus, TIGERISS will make the first definitive measurements of Ultra-Heavy Galactic Cosmic Rays (UHGCRs) on an individual element basis past barium ($^{56}$Ba), building upon the work of TIGER \cite{2009ApJ...697.2083R}, SuperTIGER \cite{2014ApJ...788...18B,Walsh:2023JH}, and CALET \cite{2025ApJ...988..148A}. These data provide a measure of the source(s) of galactic nucleosynthesis of the heaviest elements via the r-process, and these measurements have been identified as critically needed in the last Decadal Survey on Astronomy and Astrophysics (Astro2020) \cite{2021pdaa.book.....N}: “Are neutron star mergers the main site of r-process nucleosynthesis, or are there supernovae and other core collapse events that contribute significantly to the r-process budget of the universe?” and
scientists “may be able to use heavy cosmic-ray abundance measurements in the Milky Way to constrain sites of r-process nucleosynthesis.”

\section{TIGERISS Instrument}

\begin{figure}
  \centering
  \begin{minipage}{0.45\textwidth}
  \centering
  \includegraphics[width=1.0\textwidth]{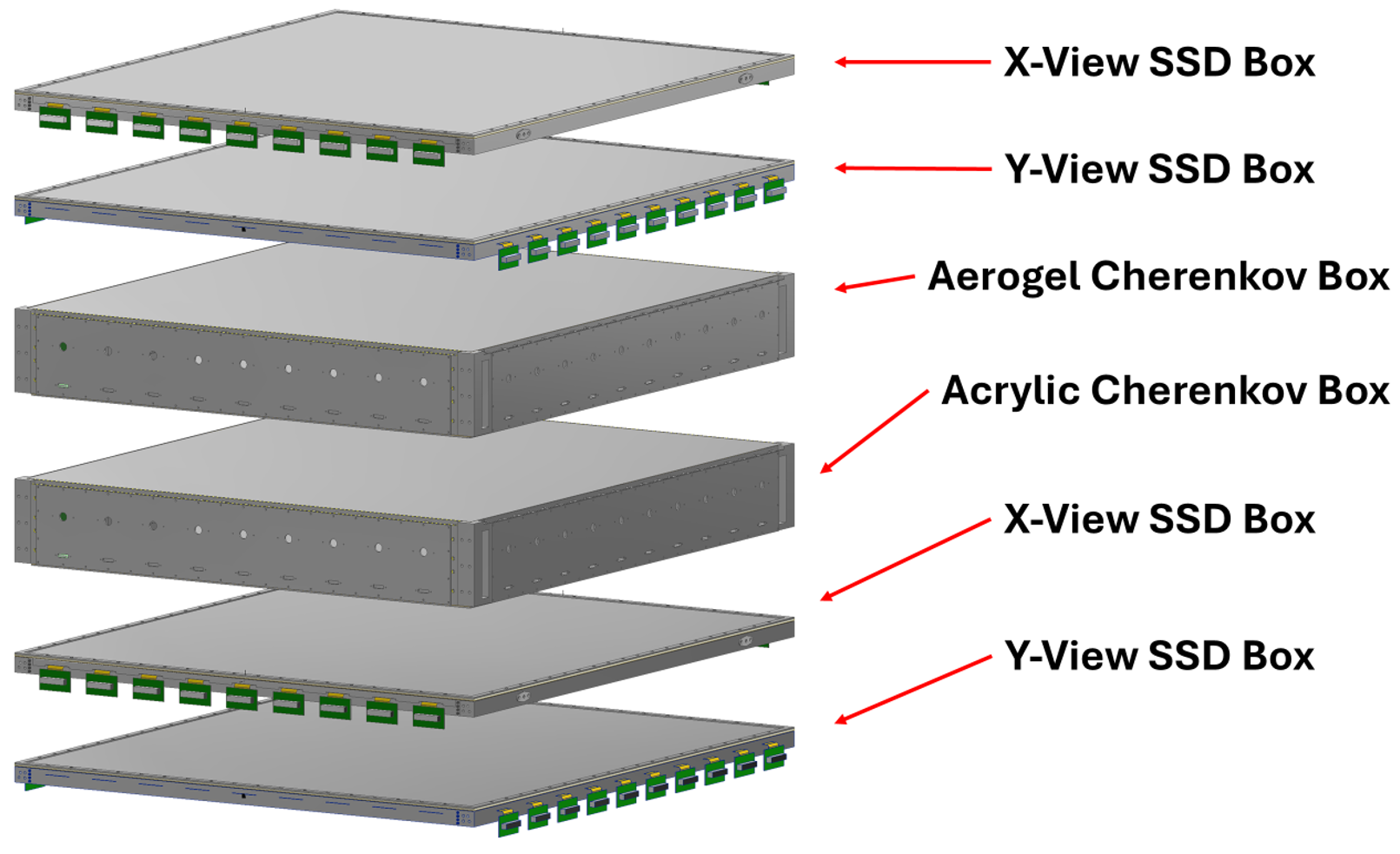}
  \caption{Expanded view of the standard TIGERISS instrument technical illustrative model.} 
  \label{fig:SSDexpd}
    \end{minipage}%
    \hspace{0.05\textwidth}
\begin{minipage}{0.48\textwidth}
  \centering
  \includegraphics[width=1.0\textwidth]{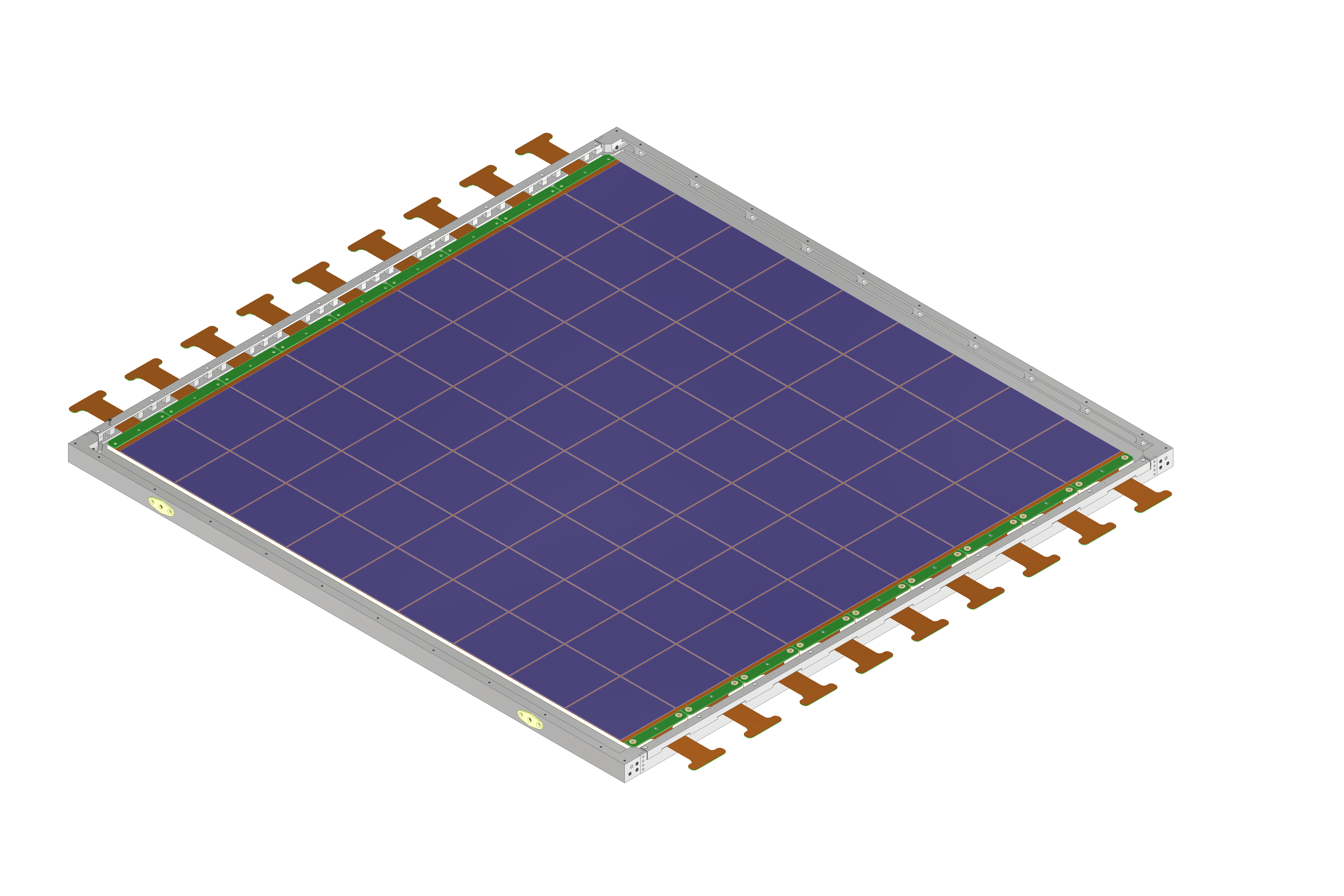}
  \caption{A TIGERISS SSD plane: Nine 9-SSD long ladder in their box.} 
  \label{fig:SSDplane}
  \end{minipage}
\end{figure}

Fig.~\ref{fig:TIGERISS} shows an illustration of the TIGERISS   payload \cite{ICRC2025tigeriss} and where it will be berthed at the SOX attachment point of the Columbus module on the International Space Station. Tab.~\ref{InstrumentTable} details the performance parameters of the TIGERISS payload. An expanded view of the TIGERISS science instrument is illustrated in Fig.~\ref{fig:SSDexpd}. TIGERISS employs four planes of single-sided silicon strip detectors (SSDs) arranged in orthogonal X-Y layers with an X-Y pair above and an X-Y pair below two large-area Cherenkov detectors. These two Cherenkov detectors measure the charge ($Z$) and velocity ($\beta$) of the traversing particles. The acrylic Cherenkov with index n=1.49, measures $\beta > 0.67$ and KE $\ge 325$ MeV/nucleon while the aerogel Cherenkov has n=1.05 and measures $\beta > 0.95$ and KE $\ge 2.12$ GeV/nucleon.

\begin{table}[h]
\begin{center}
{\small
\begin{tabular}{|l|l|} \hline
Parameter & Value  \\ \hline \hline
Payload Location &  Columbus SOX \\ \hline
Payload Mass  & 
 263 kg (CBE)\\ \hline
Payload Power &  282 W (CBE) \\ \hline
$<$Data Rate$>$ & 0.3 $-$ 0.6 Mbps \\ \hline
Maximum Data Rate & 2.4 Mbps \\ \hline
Instrument Geometry Factor & 1.06 m$^2$sr \\ \hline
Nuclei Dynamic Range & $^5$B $\rightarrow$ $^{82}$Pb \\ \hline
Charge Resolution & $<$ 0.25 charge units (cu) \\ \hline
\end{tabular}
}
\end{center}
\caption{The TIGERISS payload and instrument performance specifications.}
\label{InstrumentTable}
\end{table}

\section{Silicon Strip Detector System}

The TIGERISS silicon strip detector system is based on single-sided, DC-coupled SSDs with 100 mm $\times$ 100 mm area and 500 $\mu$m thickness characterized in CERN tests using lead and lead-fragmented beams \cite{2017ICRC...35..242K, 2019ICRC...36...94K}. These tests demonstrated measurements with single element resolution ($\sigma_{Z} < 0.25$)  from helium thru lead: $2 \lsim {\rm Z} \le 82$. This charge-resolution performance provides TIGERISS with the capability to measure nuclei at $Z \ge 56$ (barium and above) compared to that obtained with TIGER/SuperTIGER \cite{2009ApJ...697.2083R,2014ApJ...788...18B} due to high linearity of charge resolution using silicon.

\begin{figure}[!b]
    \centering
   \vspace{-2mm}
    \includegraphics[width = 0.5\textwidth]{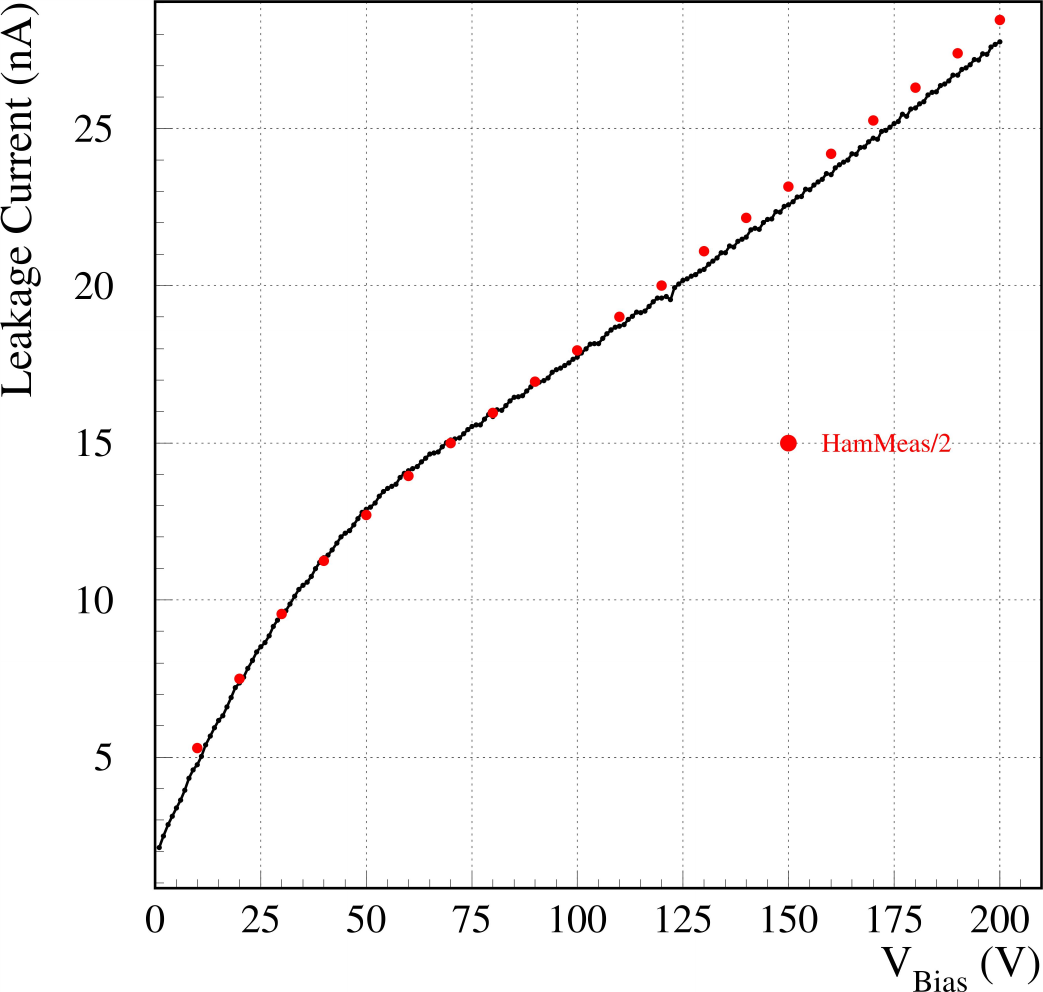}
\hspace{1mm}
\includegraphics[width = 0.48\textwidth]{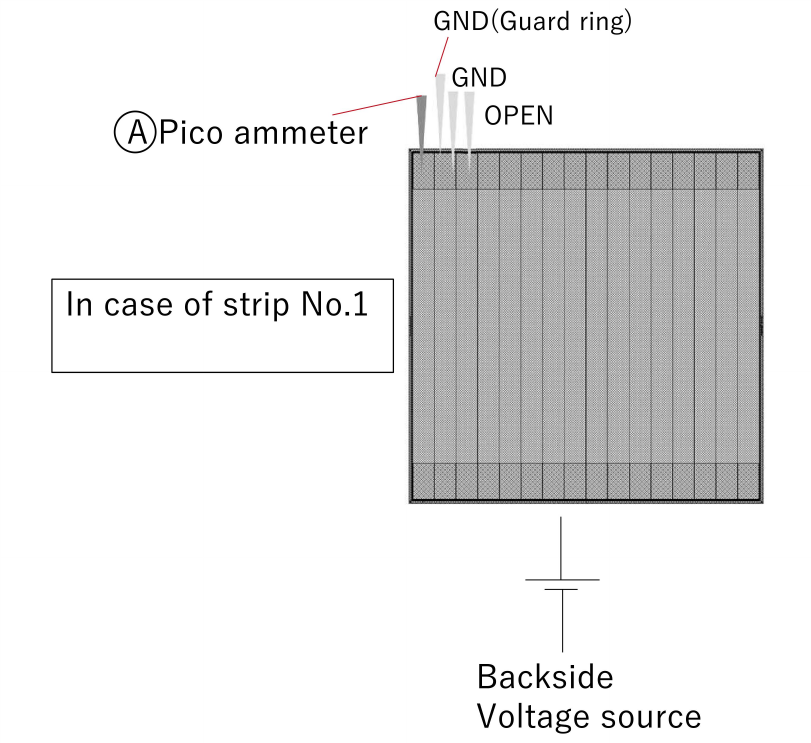}    
       \caption{Left: IV results for a strip of a Hamamatsu prototype SSD as measured at GSFC at 21$^\circ$ C (black curve) and at Hamamatsu 26$^\circ$ C (red points). The Hamamatsu values were halved to account for the different temperatures for the measurements. Right: Schematic of the IV measurement provided by Hamamatsu.}
           \label{fig:SSDtest}
\end{figure}

Tab.~\ref{SSDTable} details the performance requirements for the SSDs.  The requirement on thickness variation is needed to assure $\sigma_{Z} < 0.25$ charge resolution at the highest measured charges, eg for fully-ionized $^{82}$Pb. TIGERISS has evaluated prototype SSDs from Micron Semiconductor Ltd, Fondazione Bruno Kessler (FBK), and Hamamatsu and is procuring 400 SSDs (360 + 40 spare) from all three vendors for the TIGERISS instrument. Fig.~\ref{fig:SSDtest} presents the results of an initial measurement of a Hamamatsu prototype at the factory and at GSFC showing good agreement once the differences in temperature during the measurements were taken into account.\footnote{see S.M. Sze: Semiconductor Physics, Chapter 2 and discussion within.}

Fig.~\ref{fig:SSDplane} shows a depiction of one of the four SSD layers with the SSD ladders mounted into a mechanical frame.  Nine SSDs are daisy-chained together to form a ladder.  Nine ladders then form an SSD plane comprised of a $9 \times 9$ layout of SSDs.   The ladders are built on a mechanical substrate comprised of an aerospace-grade rohacell panel with 0.004 aluminum sheeting atached with adhesive on each side.  A custom flexcircuit provides the electrical attachment of the ohmic side of the SSDs to pads that have traces to one edge of the flexcuit to be attached to the SSD ohmic read out front-end electronics (FEE).   The SSDs are attached with adhesive and conductive epoxy to the flexcircuit which itself is adhered to the rohacell mechanical plate. The SSDS are wire bonded together then to pads on the flexcircuit that have traces to the other end of the flexcircuit for attachment to the SSD strip side FEE.  Thus, 9 channels of ohmic readout (one per SSD) is at one end of the flexcircuit while 16 channels (plus guard ring) are attached to the strip-side FEE.  The block diagrams of the FEEs are  shown in Fig.~\ref{fig:FEEblock}.

\begin{table}[h]
\begin{center}
{\small
\begin{tabular}{|l|l|} \hline
Parameter & Value  \\ \hline \hline
Physical Area & 98 mm $\times$ 98 mm \\ \hline
Active Area & 96 mm $\times$ 96 mm \\ \hline
Nominal Thickness & 500 $\pm$ 4 $\mu$m \\ \hline
Number of Strips & 16 \\ \hline
Strip Pitch & 6 mm \\ \hline
Depeletion Voltage & $\le$ 120 V \\ \hline
Total Leakage Current at +30 V above full depletion & $\le$ 500 nA \\ \hline
Strip Leakage Current at +30 V above full depletion & $\le$ 50 nA \\ \hline
Reverse Bias Breakdown Voltage & $ge$ 200 V \\ \hline
Minimum Charge Measurement & 
 $\frac{1}{2}$ Normal Incident $^5$B : 0.070 pC \\ \hline
Maximum Charge Measurement & 
 400 MeV/nuc $^83$Bi \@ 60$^\circ$ :   210 pC \\ \hline
Charge Dynamic Range &  3000 \\ \hline
Ohmic Side Detector Capacitance & $\sim$ 2000 pF \\ \hline
9-SSD long Strip Capacitance & $\sim$ 1500 pF/strip \\ \hline
\end{tabular}
}
\end{center}
\caption{The TIGERISS SSD system performance specifications.}
\label{SSDTable}
\end{table}

\subsection{Front-End Electronics}

\begin{figure}
    \begin{center}
   \vspace{-2mm}
    \includegraphics[width = 0.5\textwidth]{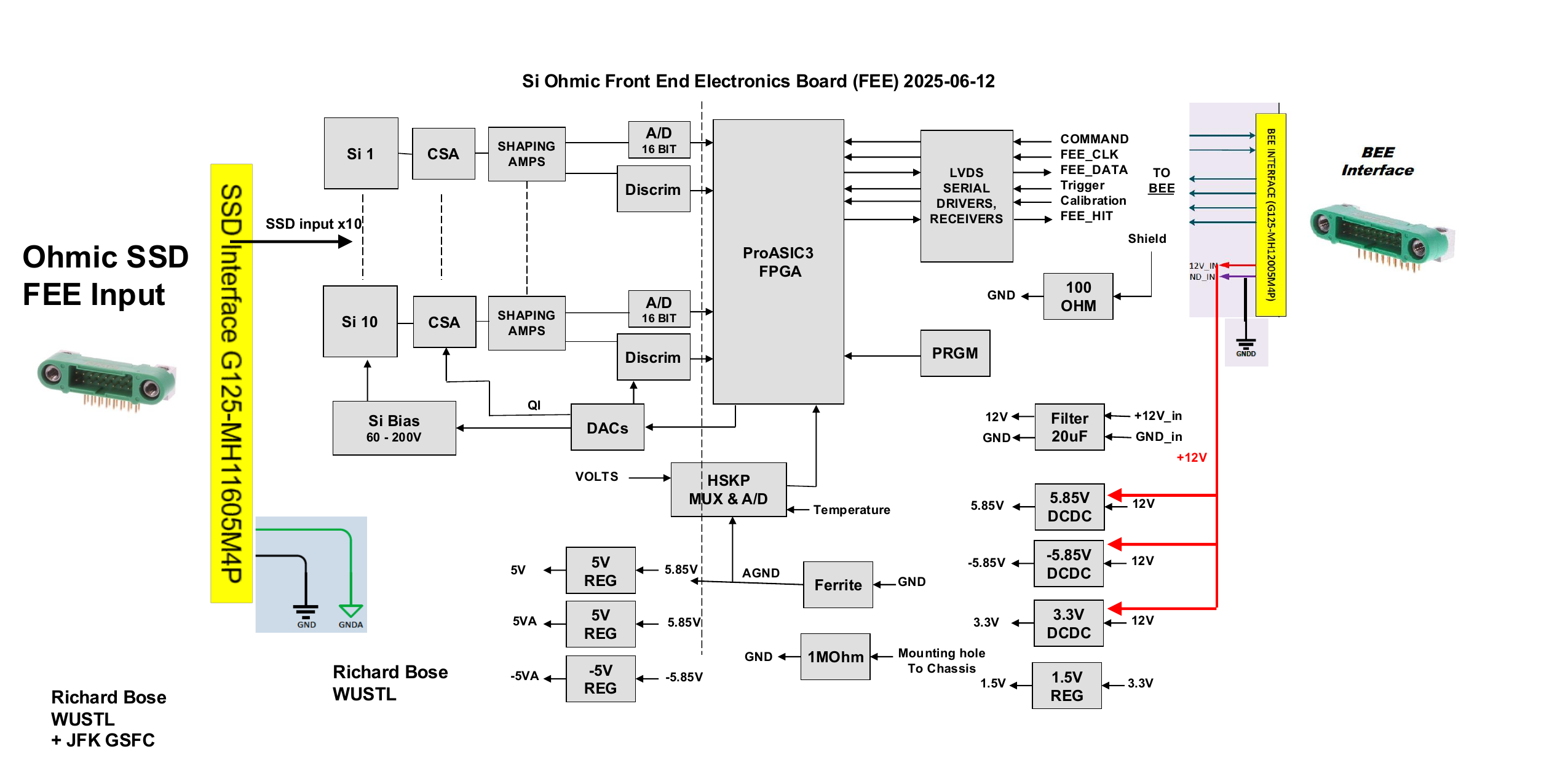}
\hspace{1mm}
\includegraphics[width = 0.48\textwidth]{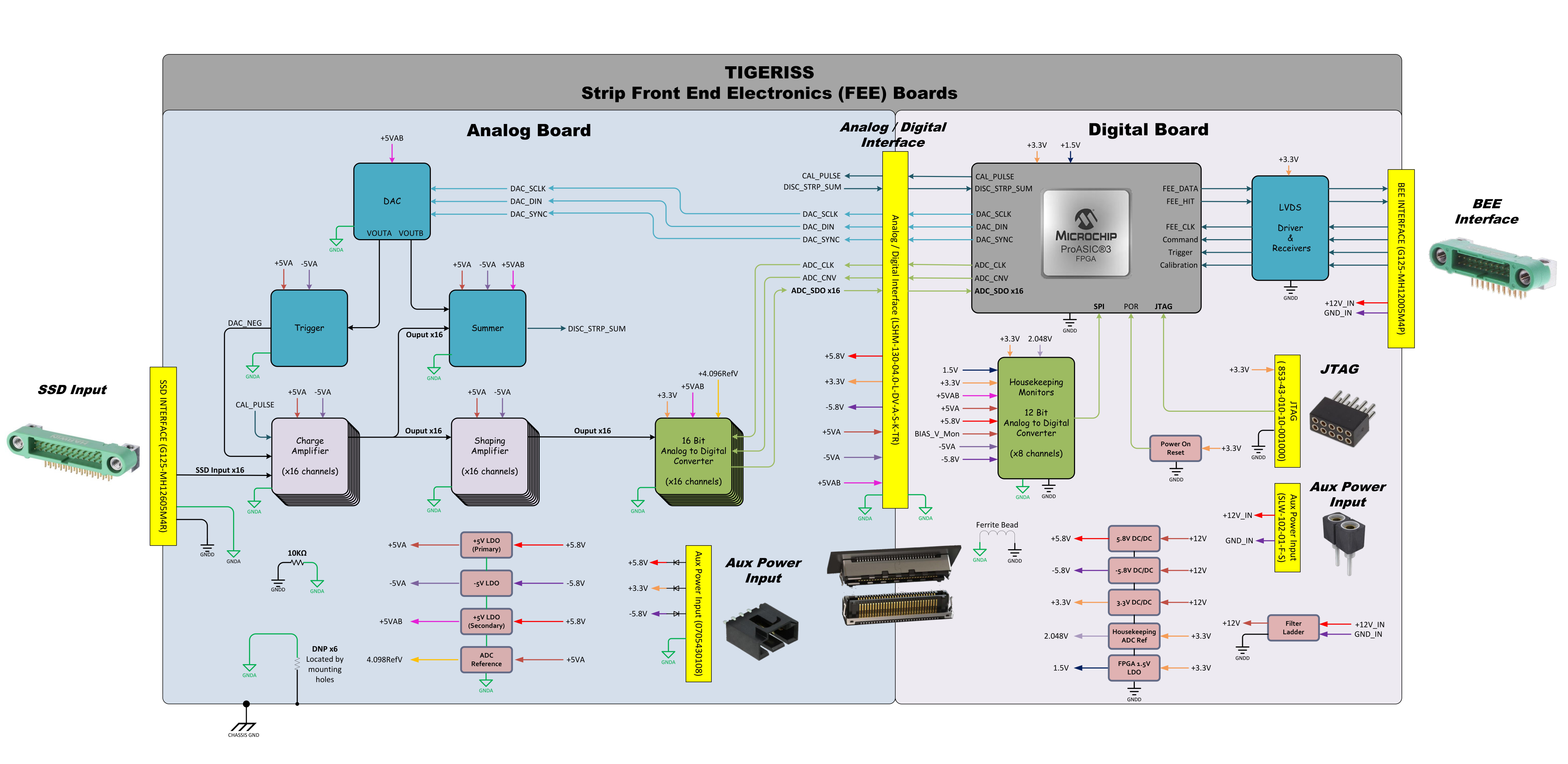}    
    \end{center}
    \vspace{-2mm}
       \caption{The block diagrams for the ohmic-side SSD front-end electronics (FEE) and the strip-side FEE.}
           \label{fig:FEEblock}
\end{figure}

Fig.~\ref{fig:FEEblock} shows the block diagram for the 9-channel ohmic FEE readout electronics and the 16-channel strip FEE electronics.  Each channel is comprised of discrete charge-shaping amplifiers, shaping amplifiers, ADS, and support electronics for charge-injection calibration. Each SSD ladder will be attached to an ohmic FEE and strip FEE on the respectives sides.   Each FEE is comprised of seperate analog and digital boards each with an onboard FPGA.  A total of 36 SSD ohmic FEE board assemblies and 36 SSD strip FEE board assemblies will be needed for the entire TIGERISS instrument. The overall instrument trigger is generated by the OR of the ohmic ionization signals from each SSD in a ladder.  A backup trigger is provided on the strip side using the OR of the strip signals on each ladder.  The analog grounds are common for the ohmic and strip FEE electronics and these are isolated from the chassis ground of the instrument. The top aluminum layer of the rohacell directly underneath each FlexCiruit is also held at the SSD FEE analog ground to minimize capacitive coupling to the mechanical structure.

\begin{figure}{t}
    \begin{center}
   \vspace{-2mm}
    \includegraphics[width = 1.0\textwidth]{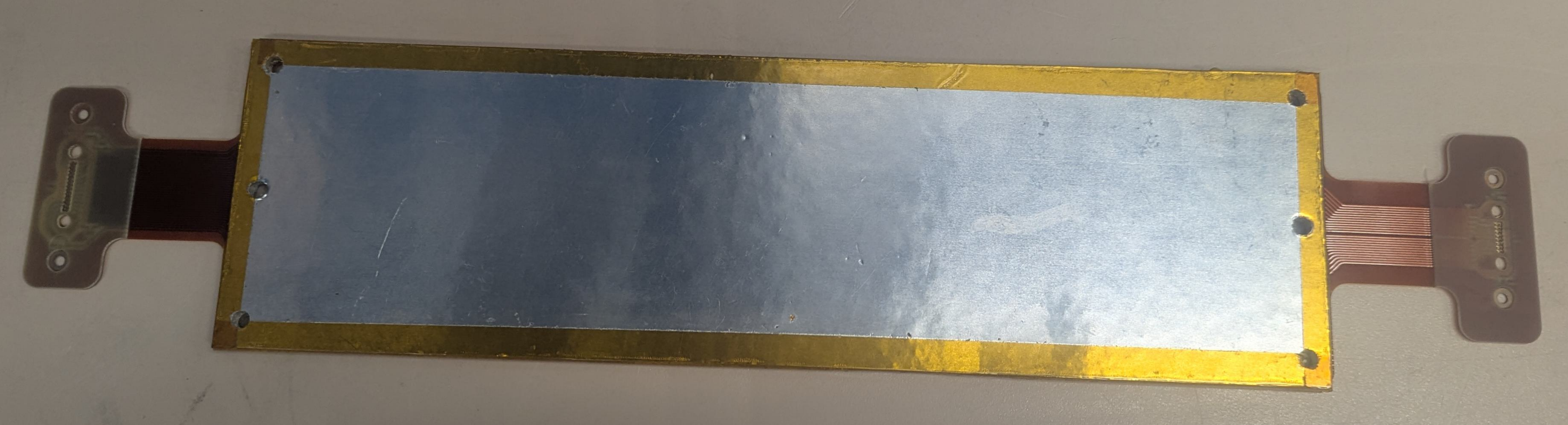}
    \vspace{2mm}
    \includegraphics[width = 1.0\textwidth]{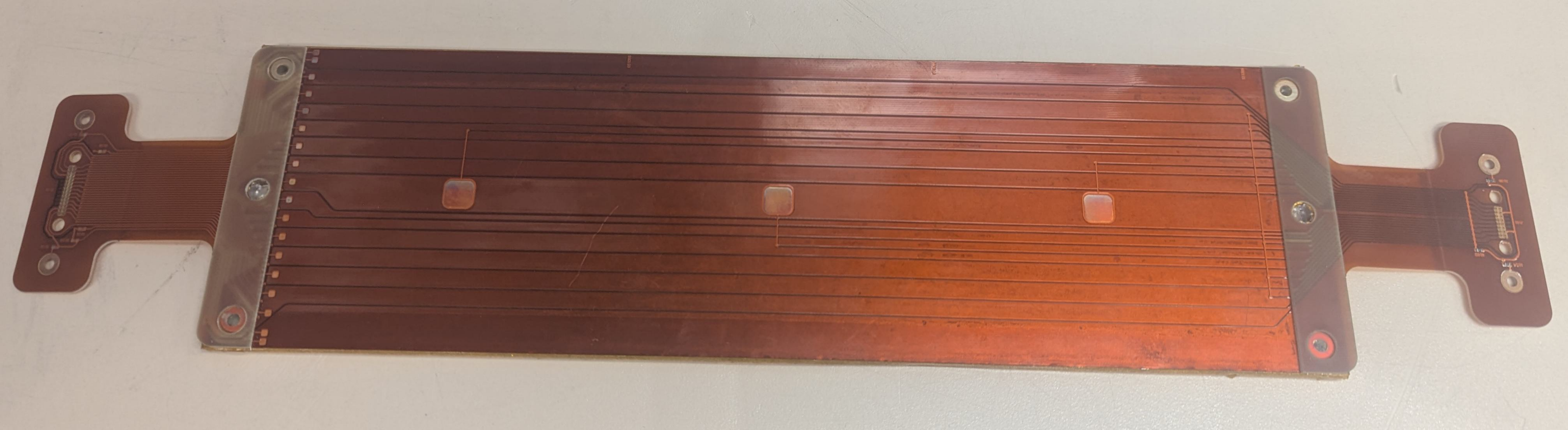}
    \vspace{2mm}
     \includegraphics[width = 1.0\textwidth]{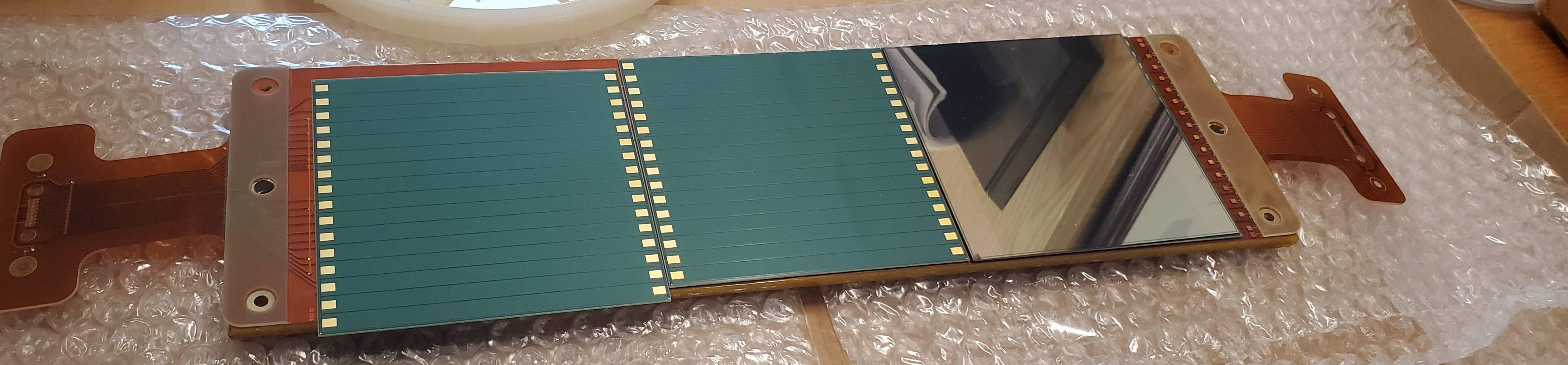}
    \vspace{2mm}
\includegraphics[width = 0.42\textwidth]{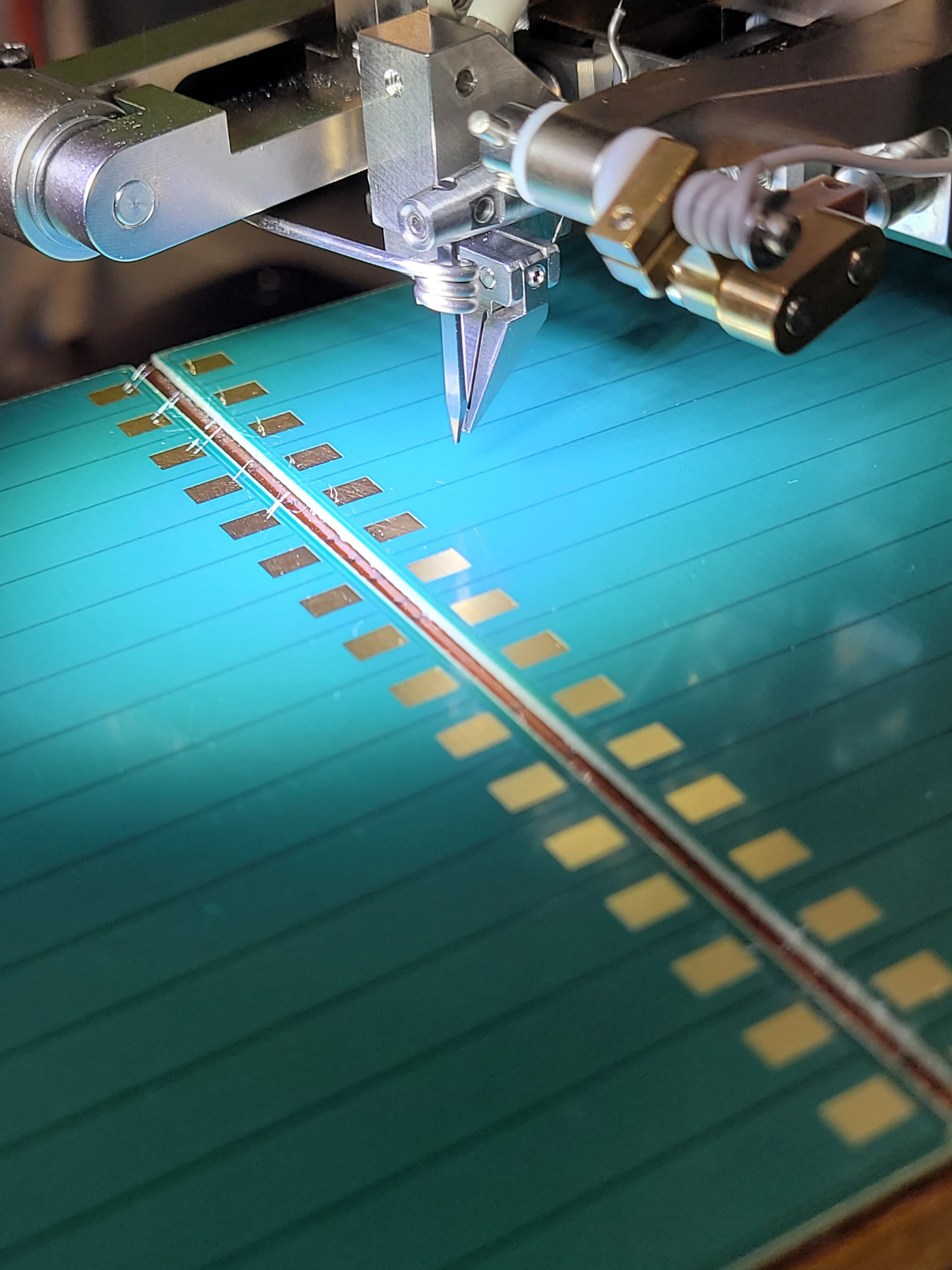}
\hspace{2mm}
\includegraphics[width = 0.48\textwidth]{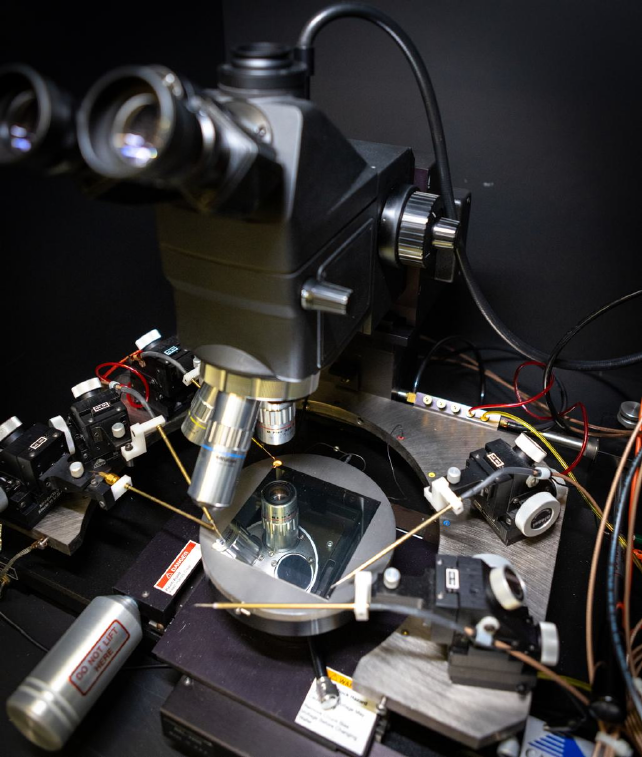}
    \caption{The process of building an Engineering Development Unit (EDU) Silicon Strip Detector (SSD) ladder (mechanical prototype using printed circuit boards vs SSDs). From top to bottom: Underside showing $\frac{1}{4}$ inch Rohacell with 4 mil aluminum sheet on each side, topside showing 3-SSD long FlexCircuit, placement of two PCBs with one Hamamatsu prototype SSD on the FlexCircuit, FlexCircuit with attached PCB dummy detectors for wire bond tests, wire bonding of the mounted PCBs, probing of a Hamamatsu prototype SSD.}

    \label{fig:EDUladder}
    \end{center}
    \vspace{-8mm}
\end{figure}

\subsection{SSD Engineering Development Unit (EDU) Ladders}

At the time of this writing, the TIGERISS development is focused on constructing Engineering Development Units (EDUs) to support near term environmental tests.  The EDUs are 3-SSD long versions of the 9-SSD long protoflight units for the full TIGERISS instrument.
Fig.~\ref{fig:EDUladder} pictorially shows the assembly process for a 3-SSD long engineering development unit (EDU). A mechanical jig is used to place the flexcircuit on the aluminum-clad rohacell (after mounting holes were machined) using transfer tape.  SSDs are then attached using conductive epoxy and transfer tape.  The strips and guard rings of the SSDs are then are wire bonded together and the last detector is wirebonded to the flexcircuit. The bottom rightmost picture in shows a Hamamatsu TIGERISS prototype detector under test with individual probes.  

\section{Acknowledgement}

This work is supported by NASA awards 21-PIONEERS21-0012 at NASA GSFC and 80NSSC22M0299 at Northern Kentucky University.

\bibliography{SSDpaper}

%

\clearpage
\section*{Full Author List: TIGERISS Collaboration}
\normalsize
%
 \noindent
H. Allen, $^{1}$ R. F. Borda, $^{2}$ R. G. Bose, $^{3}$ D. L. Braun, $^{3}$ J. Calderon, $^{4}$ Z. Campbell, $^{5}$ \\N. W. Cannady, $^{6}$ R. M.
Caputo, $^{6}$ M. Clark, $^{7}$ J. Coldsmith, $^{8}$ S. Coutu, $^{1}$ G. A. de Nolfo, $^{9}$ \\T. Forstmeier, $^{1}$ M. Fratta, $^{5}$ P. Ghosh, $^{2, 6, 10}$ S.
Graham, $^{8}$ J. F. Krizmanic, $^{6}$ \\ W. Labrador, $^{3}$ L. Lisalda, $^{3}$ J. V. Martins, $^{2}$ M. P. McPherson, $^{7}$ J. G.
Mitchell, $^{9}$ J. W. Mitchell, $^{6}$ S. I. Mognet, $^{1}$ A. Moiseev, $^{11, 6, 10}$ T. L. Ng, $^{5}$ S. Nutter, $^{4}$ N. Osborn, $^{3}$ M. Pant, $^{4}$ I. M.
Pastrana, $^{3}$ D. Radomski, $^{3}$ B. F. Rauch, $^{3}$ H. Salmani, $^{7}$ M. Sasaki, $^{11, 6, 10}$ G. E. Simburger, $^{3}$ S. Smith, $^{7}$ H. A.
Tolentino, $^{7}$ Y. Tufail, $^{5}$ D. Washington, $^{1}$ T. Widmyer, $^{5}$ L. Williams, $^{8}$ W. V. Zober, $^{3}$\\
\small\\
\noindent
$^{1}$ Pennsylvania State University\\
$^{2}$ University of Maryland, Baltimore County\\
$^{3}$ Department of Physics and McDonnell Center for the Space Sciences, Washington University in St. Louis\\
$^{4}$ Northern Kentucky University\\
$^{5}$ NASA Wallops Flight Center\\
$^{6}$ NASA Goddard Space Flight Center Astrophysics Science Division\\
$^{7}$ Howard University\\
$^{8}$ NASA Goddard Space Flight Center\\
$^{9}$ NASA Goddard Space Flight Center, Heliophysics Science Division\\
$^{10}$ Center for Research and Exploration in Space Sciences and Technology II\\
$^{11}$ University of Maryland, College Park

\end{document}